\documentclass[%
 reprint,
nofootinbib,
 amsmath,amssymb,
 aps,
pra,
]{revtex4-2}
\usepackage{graphicx}
\usepackage{dcolumn}
\usepackage{bm}
\usepackage[ruled,linesnumbered]{algorithm2e}
\usepackage{changepage}
\usepackage[labelformat=empty, position=top]{subcaption}
\usepackage[export]{adjustbox}
\usepackage{caption}
\usepackage{floatrow}
\usepackage{wrapfig}
\usepackage{float}

\usepackage{graphics}
\usepackage[T1]{fontenc}
\usepackage[utf8]{inputenc}
\usepackage{floatrow}
\usepackage[inline]{enumitem}
\usepackage{hyperref}
\usepackage{cleveref}

\usepackage{placeins}   
\usepackage{float}      

\crefname{section}{sec.}{secs.}
\Crefname{section}{Sec.}{Secs.}

\crefname{subsection}{sec.}{secs.}
\Crefname{subsection}{Sec.}{Secs.}

\crefname{equation}{Eq.}{Eqs.}
\Crefname{equation}{Equation}{Eqautions}

\crefname{figure}{Fig.}{Figure.}

\usepackage{comment}

\DeclareUnicodeCharacter{2212}{\textendash}

\DeclareMathOperator{\sech}{sech}

\newcommand{\beginsupplement}{%
        \setcounter{table}{0}
        \renewcommand{\thetable}{S\arabic{table}}%
        \setcounter{figure}{0}
        \renewcommand{\thefigure}{S\arabic{figure}}%
     }

\begin{document}

\preprint{APS/123-QED}

\title{From Chaos to Coherence: Effects of High-Order Synaptic Correlations on Neural Dynamics
}

\author{Nimrod Sherf$^{1,2}$}
\email{nsherf@uh.edu }
\author{Xaq Pitkow$^{2,3,4,5,6}$}%
\email{xaq@cmu.edu}
\author{Krešimir Josi\'{c}$^{1,7}$}%
\email{kresimir.josic@gmail.com}
\author{Kevin E. Bassler$^{1,8,9}$}%
\email{bassler@uh.edu}
\affiliation{%
$^1$Department of Mathematics, University of Houston, Houston, Texas, USA, $^2$Department of Neuroscience, Baylor College of Medicine, Houston, Texas, USA, $^3$ Neuroscience Institute, Carnegie Mellon University, Pittsburgh, Pennsylvania, USA, $^4$ Department of Machine Learning in the School of Computer Science, Carnegie Mellon University, Pittsburgh, Pennsylvania, USA, $^5$Departments of Electrical and Computer Engineering, and Computer Science, Rice University, Houston, Texas, USA, $^6$NSF AI Institute for Artificial and Natural Intelligence, $^7$Department of Biology and Biochemistry, University of Houston, Houston, Texas, USA, $^8$Department of Physics, University of Houston, Houston, Texas, USA, $^9$ Texas Center for Superconductivity, University of Houston, Houston, Texas, USA.\\
}%




\date{\today}

\begin{abstract}
Recurrent Neural Network models have elucidated the interplay between structure and dynamics in biological neural networks, particularly the emergence of irregular and rhythmic activities in cortex. However,  most studies have focused on networks with random or simple connectivity structures. 
Experimental observations find that high-order cortical connectivity patterns affect the temporal patterns of network activity, but a theory that relates such complex structure to network dynamics has yet to be developed.
Here, we show that third- and higher-order cyclic correlations in synaptic connectivities greatly impact neuronal dynamics.
Specifically, strong cyclic correlations in a network suppress chaotic dynamics and promote oscillatory or fixed activity.
This change in dynamics is related to the form of the unstable eigenvalues of the random connectivity matrix. 
A phase transition from chaotic to fixed or oscillatory activity coincides with the development of a cusp at the leading edge of the eigenvalue support. 
We also relate the dimensions of activity to the network structure.

\end{abstract}
\keywords{}
\maketitle


A central goal of theoretical neuroscience is to relate the structure, dynamics, and function of biological neural networks~\cite{sompolinsky1988chaos,sommers1988spectrum,van1996chaos,rajan2006eigenvalue,rajan2010stimulus,kadmon2015transition,aljadeff2015transition,litwin2012slow}. 
Model networks with random,  Erdős–Rényi connectivity have been studied extensively, and exhibit a range of dynamical behaviors, from fixed points and periodic orbits to highly chaotic states~\cite{sompolinsky1988chaos}. However, biological neuronal networks have highly structured connectivity \cite{white1986structure,sporns2004motifs,song2005highly,bullmore2009complex,ko2011functional,levy2012spatial,duclos2021brain,yang2023cyclic}. 
While much effort has
been devoted to understanding how such structure impacts neural dynamics and function, most work has focused on the impact of pairwise correlations, or structures encompassing the entire network~\cite{ko2011functional,rosenbaum2017spatial,rao2019dynamics,berlemont2022glassy,clark2023dimension,mastrogiuseppe2018linking,marti2018correlations}. 

\begin{figure*}[t]\hypertarget{Figure1}{}
    \centering
    \includegraphics[width=\linewidth]{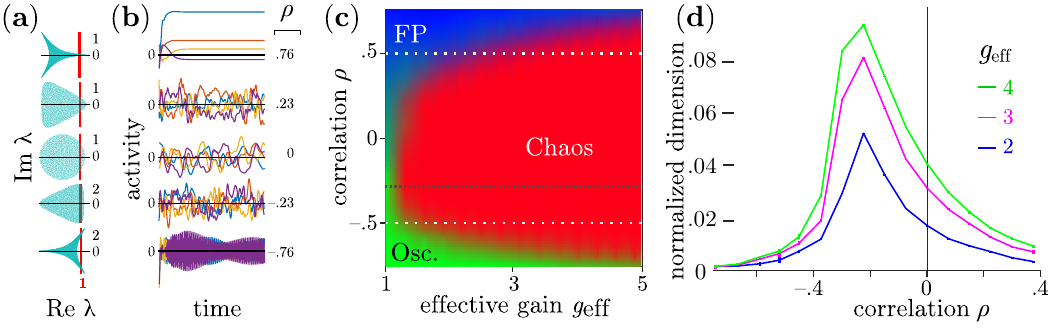}
    \caption{Dynamics of networks with third-order cyclic correlations $(\alpha=3)$: ({\bf a}) Distributions of connectivity matrix eigenvalues, $\lambda,$ at different correlation strengths, $\rho,$ for effective gain $g_{\text{eff}}=1.25$. The red lines correspond to Re $\lambda=1$; eigenvalues to the right of the red line are unstable. ({\bf b}) Activities of four typical neurons at each value of $\rho$. ({\bf c}) Heat map showing probabilities of observing chaotic, oscillatory, or fixed point activity (red, green, and blue, respectively) as a function of $g_{\text{eff}}$ and $\rho$. White dashed lines correspond to $\rho=\pm0.5$. The black dotted line at $\rho \approx -0.28$ depicts the value at which the probability of observing chaotic behavior was highest when $g_{\text{eff}}=1.2$. ({\bf d}) Averaged normalized attractor dimension (participation ratio) at constant values of $g_{\text{eff}}$ as a function of $\rho$.}
    \label{fig:fig1}
\end{figure*} 

Mounting evidence suggests that cortical architecture is characterized by high-order correlations between synaptic connections~\cite{white1986structure,song2005highly,perin2011synaptic,sporns2011non,duclos2021brain,wei2017identifying,yang2023cyclic}. Motifs composed of three or more neurons are over-represented, and shape neural activity \cite{pernice2011structure,hu2013motif,hu2014local,gollo2014frustrated,gollo2015dwelling,wei2017identifying,battiston2021physics,deshpande2023third}, yet the impact of such structure on neural dynamics and function remains poorly understood. While progress has been hindered by limits on our understanding of networks with higher order correlations in connectivity, advances in random matrix theory \cite{metz2011spectra,bolle2013spectra,ahmadian2015properties,kuczala2016eigenvalue,aceituno2019universal}, and statistical physics \cite{helias2020statistical,zou2023introduction,clark2023dimension} are paving the way for new insights.

Here, using numerical simulations we show that third- and higher-order cyclic correlations in synaptic connectivities have a strong impact on neuronal dynamics. 
We then characterize the connectivity structure through its eigenvalue spectra and relate it to the network dynamics. We find that the emergence of irregular activity is strongly affected by high-order correlations in connectivity: Strong, positive high-order correlations tend to stabilize neuronal dynamics and reduce the dimensionality of activity. In contrast, strong, negative high-order correlations lead to high-frequency rhythmic behavior. Our results suggest that networks with such connectivity structure do not display chaotic behavior, unlike unstructured \cite{sompolinsky1988chaos} and partially symmetric networks \cite{marti2018correlations}.

We consider a rate model of neuronal activity in a network of $N$ neurons \cite{amari1972characteristics,sompolinsky1988chaos,mastrogiuseppe2018linking,clark2023dimension}, 
\begin{equation}  \label{eq:dyn}
\dot{x}_i(t)=-x_i(t)+\sum_{j=1}^{N}w_{ij}\phi(x_j(t)), \qquad  i = 1, \ldots, N,
\end{equation}
where $x_i$ is the membrane potential of neuron $i$, $\phi(x_j)=\tanh(x_j)$ is the activation function of neuron $j$, and $w_{ij}$ is the weight of the synaptic connection from neuron $j$ to neuron $i$.
Weights satisfy
$\langle w_{ij} \rangle=0,   \langle w^2_{ij} \rangle =g^2/N, $ and
\begin{equation}\label{eq:w_stat}     
\underbrace{ \langle w_{ij} w_{jk} w_{kl} ... w_{pi}  \rangle}_{\alpha}    =\frac{g^{\alpha} \rho}{N^{\alpha-1}}  
\end{equation}
where the angular brackets $ \langle ... \rangle$ denote the ensemble average. 
The gain parameter $g$ determines the coupling strength, 
$\rho$ controls the strength of high-order directed cyclic correlations, and $\alpha$ denotes correlation order. We used a modification of the algorithm proposed by Aceituno, et al. \cite{aceituno2019universal} to generate matrices with the desired statistics (see \Cref{supp:matrix_generation} of supplementary materials (SM) for details).

\begin{figure*}[t]\hypertarget{Figure2}{}
    \centering
        \includegraphics[width=\linewidth]{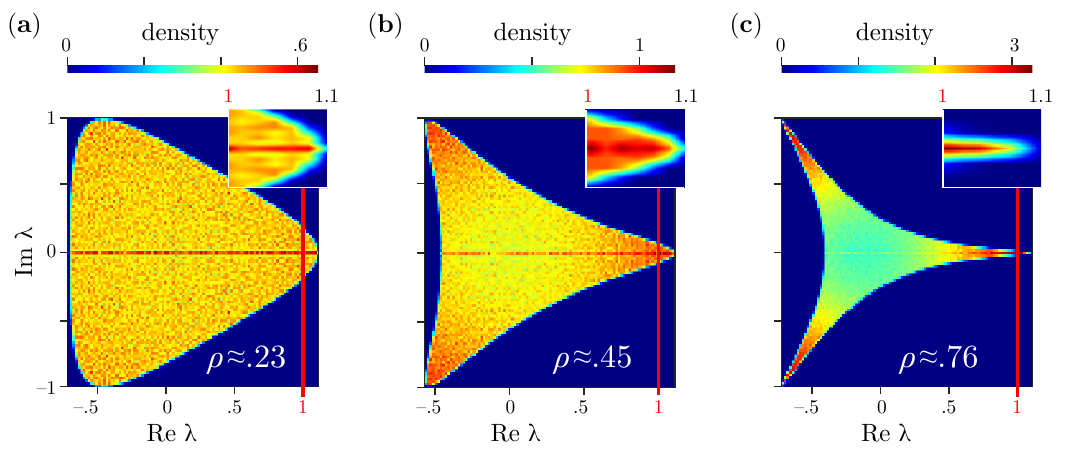}
    \caption{Distribution of  connectivity matrix eigenvalues when $\alpha = 3, g_{\text{eff}}=1.1,$ and ({\bf a}) $\rho \approx 0.23$, ({\bf b})  $\rho \approx 0.45$, and ({\bf c})  $\rho \approx 0.76$. Red lines at $\mbox{Re}\; \lambda=1$ mark the limit of stability for eigenvalues. Insets show expanded views of the distributions of the unstable eigenvalues at the tips that control the network dynamics. }
    \label{fig:fig2}
\end{figure*}

The eigenvalue spectrum of random matrices with high-order cyclic correlations obeys a Hypotrochoid Law~\cite{metz2011spectra,bolle2013spectra,aceituno2019universal}: The eigenvalue density when
$|\rho| \ll 1$ and $N \rightarrow \infty$ has support inside the hypotrochoid $z(\phi)=g(e^{i \phi}+\rho e^{-i(\alpha-1) \phi})$.  
In contrast to uncorrelated or pairwise correlated weights, the eigenvalue density when $\alpha \geq 3$ is not uniform within the hypotrochoid, ~\cite{ginibre1965statistical,girko1985circular,tao2010random,aceituno2019universal}.
At finite values of $\rho$, we observe deviations from the Hypotrochoid Law (see \hyperlink{FigureS2}{Fig.\ S2 of SM}): At small values of $|\rho|$, eigenvalues can be found just outside the hypotrochoid, and at $|\rho|>\rho_{\text{c}}$, where $\rho_{\text{c}} \equiv (\alpha-1)^{-1}$, hypotrochoids have loops, which the eigenvalue density does not follow.
 
The Hypotrochoid Law, nevertheless, provides an approximation of the real part of the leading (rightmost) eigenvalue, which we call the effective gain, $g_{\text{eff}}$. In general, the phase of the hypotrochoid that corresponds to the leading eigenvalue is a solution of $U_{\alpha-2}(\cos(\phi^*))=-(\rho(\alpha-1))^{-1}$, where $U_{\alpha}(x)$ are the Chebyshev polynomials of the second kind. At
\begin{equation}\label{eq:rhof}
\rho_{\text{f}}\equiv-(\alpha-1)^{-2},
\end{equation}
the leading edge of the eigenvalue distribution is on the real axis, and the curvature of the support at that point is zero. For $\rho \geq\rho_{\text{f}}$, the leading edge of the eigenvalue support has a unique solution corresponding to $\phi^*=0$, while for $\rho<\rho_{\text{f}}$ it has two solutions. The effective gain, $g_{\text{eff}},$ is the real part of the support at $\phi^*$, and for $\alpha=3$,
\begin{equation}\label{eq:geffold}
   g_{\text{eff}} \equiv   \left\{
    \begin{array}{ll}
         g(1+\rho) & \rho \geq \rho_{\text{f}}= -0.25 \\
        -g(\rho+1/(8\rho))  & \rho <\rho_{\text{f}}= -0.25. 
    \end{array}
    \right.
\end{equation}
When $g_{\text{eff}} \lesssim
1$ the fixed point  $\mathbf{x}=\mathbf{0}$ is stable.
This ``quiescent'' fixed point becomes unstable when the real part of the leading eigenvalue is greater than $1$, so the onset of nontrivial dynamics occurs at $g_{\text{eff}}\approx1$.
In large networks with uncorrelated weights, high-dimensional chaotic activity emerges as soon as the quiescent fixed point becomes unstable~\cite{sompolinsky1988chaos}.

In contrast to networks with uncorrelated weights, we find that high-order cyclic correlations ($\alpha \geq 3$)  suppress the emergence of chaotic activity near the onset of nontrivial dynamics.
\hyperlink{Figure1}{Figs.\ 1(a)--(b)}
show representative network activity near this onset. When weights are uncorrelated, $\rho = 0$, chaotic activity emerges near the onset of instability of the origin,  
and persists when correlations are weak and positive. However, solutions typically converge to a fixed point when correlations are strong ($\rho \approx 0.76$ in \hyperlink{Figure1}{Fig.\ 1(b)}). Similarly, chaotic activity persists when correlations are weak and negative, $\rho \approx -0.23$, but is replaced by high frequency oscillations when correlations are strong and negative, $\rho \approx -0.76$.

\hyperlink{Figure1}{Figure\ 1(c)} shows the probability of observing chaotic, oscillatory, or fixed point dynamics as $g_{\text{eff}}$ and $\rho$ are varied with the intensity of the different colors representing the empirically determined probability of each state. To obtain these probabilities we smoothed the results from 300 simulations at each point of a parameter grid with spacings $\Delta g_{\text{eff}}\approx 0.29$ and  $\Delta \rho \approx 0.08$, with each simulation using a different network realization (see \Cref{supp:numerical_methods} of SM for details). 
 The fact that oscillations or fixed points are sometimes found near the onset of instability of the quiescent fixed point when $|\rho|<\rho_{\text{c}}$ is presumably a finite size effect, as is known analytically to be the case for $\rho=0$ \cite{sompolinsky1988chaos}. In \hyperlink{Figure1}{Fig.\ 1}, the results are shown for $N=1600$.
 The range of $g_{\text{eff}}$ for which fixed points or oscillations occur after this onset decreases in width as $1/\sqrt{N}$ for $\rho=0$ \cite{sompolinsky1988chaos}. We find that this range also decreases with $N$ for $|\rho|<\rho_{\text{c}}$ (see \hyperlink{FigureS3}{Fig.\ S3(a)-S3(b) of SM}). Numerical scaling shows that non-chaotic behavior vanishes in the limit of $N\rightarrow\infty$ as $g_{\mathrm{eff}} \rightarrow 1$ and $|\rho|<\rho_{\text{c}}$ (see \hyperlink{FigureS4}{Fig.\ S4(a)-S4(b) of SM}). In contrast, when $|\rho|>\rho_{\text{c}}$, near the onset we still frequently observed non-chaotic solutions numerically  even when $N$ was large (see \hyperlink{FigureS3}{Fig.\ S3(c)} of SM). However, even for $|\rho|>\rho_{\text{c}}$, chaotic solutions were detected far from the onset, at large values of $g_{\mathrm{eff}}>1$  (see \hyperlink{FigureS4}{Fig.\ S4(c) of SM}). Although in this regime, transient effects can cause an overestimation of the probabilities of observing chaotic dynamics (see \Cref{supp:Fitting chaos probabilities} of SM). 
 

In recurrent neural networks, the dimensionality of attractors is constrained by the network architecture~\cite{clark2023dimension}. The dimensions of the sub-space spanned by these attractors can affect a network's ability to learn~\cite{sadtler2014neural}, and reflect properties of external stimuli \cite{chaudhuri2019intrinsic}. A common measure of dimensionality is the participation ratio normalized by network size~\cite{recanatesi2019dimensionality,recanatesi2022scale,clark2023dimension}. \hyperlink{Figure1}{Fig.\ 1(d)} shows that this normalized participation dimension peaks
 at $\rho \approx -0.3$ for different values of $g_{\text{eff}}$,
 and decreases considerably from this peak at other values of $\rho$.  Thus, even when solutions are chaotic, third order cyclic correlations strongly impact network dynamics.

What causes the observed stabilization of network dynamics? We hypothesize that this is due to the distribution of unstable eigenvalues, \emph{i.e.} eigenvalues $\lambda$ with $\mbox{Re}\; \lambda>1$.
The shape of the support of the eigenvalue spectrum, approximately described by the hypotrochoid curve, changes with $\rho$ (See \hyperlink{Figure1}{Fig.\ 1(a)} for $\alpha=3$).
In general, the support has $\alpha$ ``vertices''. The eigenvalue distributions at $\rho$ and $-\rho$ are identical up to a rotation by $\pi/\alpha$.  
 For $\rho>\rho_{\text{f}}$ (see \cref{eq:rhof}), the leading edge is on the real axis, while for $\rho<\rho_{\text{f}}$, the leading edge occurs at two points with the real axis half way between them. 
When $|\rho|<\rho_{\text{c}}$, the eigenvalue support is smooth at the vertices, but when $|\rho|\geq\rho_{\text{c}}$, the support has cusps. We observed numerically that this change in the morphology coincides with a phase transition in the dynamics at the onset of instability of the quiescent fixed point. Thus, $\rho=\pm \rho_{\text{c}}$ are critical points.

This morphological change as a function $\rho$ also involves the development of a nonuniform eigenvalue density as shown in \hyperlink{Figure2}{Fig.\ 2}, which is calculated for $\alpha=3$ ($100$ realizations, $N=6400$).
In particular, eigenvalues accumulate near the tips of the vertices as $|\rho|$ increases, with an excess of real eigenvalues observed for all values of $\rho$. For networks with pairwise correlations ($\alpha=2$), this excess of real values is known to decay as $\sim 1/\sqrt{N}$, and thus disappears as the network grows in size \cite{sommers1988spectrum}. 
For weak correlations, the eigenvalue distribution is nearly uniform, e.g., at $\rho \approx 0.23$ as shown in \hyperlink{Figure2}{Fig.\ 2(a)}. As $\rho$ increases, the eigenvalues accumulate near the tips of the vertices, e.g., at $\rho \approx 0.45$ as shown in \hyperlink{Figure2}{Fig.\ 2(b)}. However, for $\rho>\rho_{\text{c}}=0.5$,
when the vertices are cusps, the eigenvalue density is maximal at a point in the middle of each cusp, away from the tips. Here, the eigenvalue density decays slowly towards the tips from the maxima, as can be seen for $\rho \approx 0.76$ in \hyperlink{Figure2}{Fig.\ 2(c)}.

The leading eigenvalues in the spectrum control the stability of the dynamics. For $\rho=0$, the spectrum is given by the circular law \cite{sompolinsky1988chaos,tao2008random}, and the leading eigenvalues are at the leading edge of the circle.
In the limit of large $N$, the real part of many eigenvalues simultaneously exceeds $1$  at the onset of instability of the quiescent fixed point where $g_{\text{eff}}\approx 1^+$, leading to chaos ~\cite{sompolinsky1988chaos}. The same appears to be true for all $|\rho|<\rho_{\text{c}}$: the real part of many eigenvalues simultaneously exceeds $1$  and chaotic dynamics emerge with high probability.
For $\rho\approx \rho_{\text{f}}$, where the leading edge of the spectrum is nearly flat (vertical), dynamics has maximal dimensionality (see \hyperlink{Figure1}{Fig.\ 1(d)}). At this value of $\rho$, the probability of observing chaotic dynamics when the origin becomes unstable is also maximal, as shown in \hyperlink{Figure1}{Fig.\ 1(c)} (see also \hyperlink{FigureS5}{Fig.\ S5 of SM}.)

As $\rho$ increases for $\rho>\rho_{\text{f}}$, the leading edge of the eigenvalue distribution becomes increasingly pointed and the magnitudes of the imaginary parts of the eigenvalues decrease, thus the fluctuations in the dynamics slow. (See insets of \hyperlink{Figure2}{Figs.\ 2(a) \text{and} 2(b)}.) 
As $\rho$ decreases for $\rho<\rho_{\text{f}}$, the eigenvalues near the leading edge accumulate near the two imaginary points, leading to stable oscillations near the onset of instability of the quiescent fixed point. In both cases, for $|\rho|>\rho_{\text{c}}$, when there are cusps at the tips of the vertices, eigenvalues do not accumulate at the edge of the support. Instead, the eigenvalue density decays toward the tips of the cusps.
Because of these properties, the number of unstable eigenvalues increases slowly with $g_{\text{eff}}$ after the onset, even in the limit of large $N$.  (See inset of \hyperlink{Figure2}{Fig.\ 2(c)}.) This slow accumulation of eigenvalues corresponds with the high probability of observing fixed points or oscillations in the dynamics. 
Thus, at $\rho=\pm\rho_{\text{c}}$, where there is a morphological change at the leading edge of the eigenvalue support, we conjecture that there are critical points where phase transitions in the dynamics at onset occur.
This conclusion is consistent with what is known about networks with pairwise correlations ($\alpha=2$), where chaos occurs at onset for all $|\rho|<1$, but trajectories converge to equilibria at $\rho=\rho_{\text{c}}=1$ \cite{marti2018correlations,crisanti2018path}.

Although chaotic dynamics do not occur at the onset of instability of the origin for strong correlations where $|\rho|>\rho_{\text{c}}$, chaotic dynamics does emerge when $g_{\text{eff}}$ is large, e.g., when $|\rho| \approx 0.75$ chaotic behavior is still observed with high probability when $g_{\text{eff}} \gtrsim 3$ in \hyperlink{Figure1}{Fig.\ 1(c)}.
The value of $g_{\text{eff}}$ for which chaotic dynamics is first observed decreases as $|\rho|\rightarrow \rho_{\text{c}}$, presumably to $\sim 1^+$ in the limit of large $N$. 
However, simulations are delicate in this region because the transient time increases with both $g_{\text{eff}}$ and $N$, and it becomes difficult to distinguish between chaotic behavior, and fixed point or oscillatory dynamics with long transients.

 Numerical investigations of networks with higher order correlations, $\alpha=4, 5\ \text{and} \ 6$ confirm the existence of phase transitions in the dynamics at the onset of instability of the fixed point at the origin and critical points at $\rho=\pm \rho_\text{c}$ consistent with the observation in the case $\alpha = 3$.
However,  at values of $|\rho| \gg \rho_{\text{c}}$, deviations from the Hypotrochoid Law become increasingly large with $\alpha$.

\begin{figure}\hypertarget{Figure3}{}
    \centering    \includegraphics[width=\linewidth]{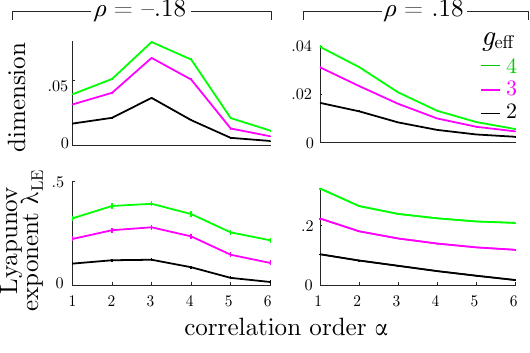}

    \caption{Averaged normalized attractor dimension (top row) and corresponding Lyapunov exponents, $\lambda_{\text{LE}}$, (bottom row) as a function of $\alpha$ for different values of $g_{\text{eff}}$. ($\alpha=1$ corresponds to uncorrelated networks.) The left and right columns show results with negative and positive correlations $\rho \approx \pm 0.18$, respectively.}
\end{figure}

Dynamics in the region where chaotic behavior is observed are strongly influenced by high-order correlations. \hyperlink{Figure3}{Figure\ 3} shows the average attractor dimensionality and Lyapunov exponents, $\lambda_{\text{LE}},$ as functions of correlation order, $\alpha$, at fixed values of $|\rho| \approx 0.18$ and $g_{\text{eff}}$, obtained by averaging over $50$ realizations of networks with $N=1600$. (See  \hyperlink{FigureS6}{Fig.\ S6 of SM} for the corresponding representative eigenvalue distributions.) Dimensionality and Lyapunov exponents behave similarly:
For positive values of $\rho$, these two quantities decrease monotonically with $\alpha$. For negative values of $\rho$, both dimensionality and Lyapunov exponents grow as the spectrum of eigenvalues flattens at the leading edge of the spectrum on the real axis, reaching a peak when $\rho=\rho_\text{f}$. At the correlation value $\rho_\text{f}$ the spectrum at this leading edge is the flattest. This value depends on the correlation {\it order,} $\alpha$, and for third-order correlations the spectrum is flattest at $\rho=-.18$. This is consistent with the value of $\rho$ where we observed that dimensionality and Lyapunov exponents reach their maximum values, as shown in  \hyperlink{Figure3}{Fig.\ 3}. Similarly, dimensionality as a function of correlation also peaks at $\rho \approx \rho_{\text{f}}$ as shown in \hyperlink{Figure1}{Fig.\ 1(d)}.

Thus, we conjecture the phase diagram in \hyperlink{Figure4}{Fig.\ 4} for the dynamics at the onset of instability of the quiescent fixed point as a function of $\rho$ and $\alpha$. 
Lines of critical points (solid lines) separate the chaotic phase (red) from the phases where fixed points (blue) and stable oscillations (green) occur. The dimensionality of the dynamics and the Lyapunov exponents are maximal along a line in the middle of the chaotic region (dashed line).

\begin{figure}[ht]\hypertarget{Figure4}{}
    \centering
    \includegraphics[width=0.55\textwidth]{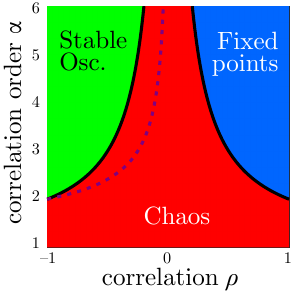}
    \caption{Phase diagram of dynamics at the onset of the instability of the origin as a function of correlation strength $\rho$ and correlation order $\alpha$ in the limit of $N \rightarrow \infty$. Solid lines correspond to $\pm\rho_{\text{c}}$, where the eigenvalue support of the synaptic connectivity matrix develops cusps. The dashed line corresponds to $\rho_\text{f}$, where the leading edge of the support is nearly flat and the dimensionality of the dynamics is maximal.}
\end{figure}

Biological neural networks are known to exhibit high-order cyclic synaptic correlations \cite{song2005highly,yang2023cyclic,wei2017identifying}. We have shown how such connectivity structures influence network dynamics by examining a minimal model of cortical networks designed to capture the impact of cyclic correlations. Networks with strong positive correlations support stable or weakly chaotic dynamics, which are known to enhance computational performance \cite{bertschinger2004edge,laje2013robust,kozachkov2020achieving}. Networks with strong negative correlations support robust oscillations, which are a hallmark of neural dynamics. The strong relation between low-dimensional chaotic dynamics and high-order structures, suggests that low-dimensional neural activity may a consequence of \emph{local} structure in synaptic connectivity. These findings may also provide insights into the macroscopic statistical properties of other disordered correlated systems, such as spin glasses \cite{larson2010numerical,bagherikalhor2023frustration,mezard2024spin}, ecological networks \cite{case1981testing,bairey2016high}, and social networks \cite{battiston2020networks}.

Mounting evidence shows that network connectivity patterns shape both learning and computation \cite{mante2013context,sadtler2014neural,mastrogiuseppe2018linking,zavitz2021connectivity,schirner2023learning,chia2023emergence}.
Yet, the interplay between the full spectrum of high-order correlations and learning remains largely unexplored.
By outlining how higher-order correlations reshape network dynamics, our work provides a concrete framework for testing their computational impact. Pursuing this line of inquiry promises not only to clarify the circuit principles that govern cortical function, but also to inspire new and novel artificial learning models. In addition, embedding these correlations in network models yields biologically plausible architectures that generate specific, testable predictions for future experiments.\\

\section*{\label{sec:ACKNOWLEDGMENTS}ACKNOWLEDGMENTS}
This research was supported in part by grants from NSF NeuroNex (DBI-1707400), NIH (R01 MH130416) to NS, XP and KJ, NSF (DMS-2207647) and NSF (DMS-2235451)  to KJ, and Simons Foundation (MPS-NITMB-00005320) through the NSF-Simons National Institute for Theory and Mathematics in Biology to NS, XP, KJ, and KEB.

\bibliographystyle{apsrev4-1}

\bibliography{MainPaper}

\clearpage                        
\phantomsection

\section{\label{Supp}Supplementary Materials}
\beginsupplement
\subsection{Matrix generation}\label{supp:matrix_generation}
To generate $\alpha$-order cyclic correlations, we start with an $N \times N$ random matrix, where the entries are iid and $\sim \mathcal{N}(0,g^2/N)$. Next, we methodically change the sign of the corresponding entries such that the sum of all cycles of length $\alpha$ that end up in that specific entry would be positive with a probability $P$. This probability determines the strength of the correlations $\rho$, see \hyperlink{FigureS1}{Fig.\ S1}. We do the same for negative correlations, but demand that the sum be negative. The step-by-step description is given in \Cref{alg:corr}.  
\IncMargin{1.5em}
\begin{algorithm}[ht]
    \SetKwData{Left}{left}
    \SetKwData{This}{this}
    \SetKwData{Up}{up}
    \SetKwFunction{Union}{Union}
    \SetKwFunction{FindCompress}{FindCompress}
    \SetKwInOut{Input}{input}
    \SetKwInOut{Output}{output}
    \Input{$N \times N$ random matrix with $w_{ij}\sim \mathcal{N}(0,g^2/N)$}
    \Output{$N \times N$ matrix with $\mathrm{Tr}\,W^{\alpha}/N=\rho>0$}
    \BlankLine

    Initialize an $N \times N$ random matrix, where the entries are iid with $w_{ij}\sim \mathcal{N}(0,g^2/N)$\;
    \For{$n=\alpha-1:N$}{
        $\tilde{w}=(w_{n+1,1:n} w_{1:n,1:n}^{\alpha-2})\odot w_{n+1,1:n}$\;
        \For{$c=1:n$}{
            \If{$\tilde{w}_{c,n+1}<0$}{
                $w_{c,n+1} \rightarrow -w_{c,n+1}$ by a probability $P$\;
            }
        }
    }
    \caption{Generating high-order cyclic correlations}\label{alg:corr}
\end{algorithm}
\DecMargin{1.5em}
Note that $w_{l,1:k}$ is a $1 \times k$ vector, and $\odot$ is the element-wise product.

\begin{figure}[h]\hypertarget{FigureS1}{}
    \centering
    \includegraphics[width=0.65\textwidth]{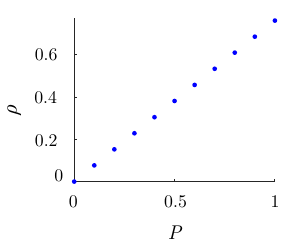}
    \caption{Numerical value of $\rho$ as a function of the probability $P$, averaged over 30 realizations.}
    \label{fig:rho_vs_P}
\end{figure}

\begin{figure}\hypertarget{FigureS2}{}
    \centering  \vspace{0.5cm}
     \begin{subfigure}[t]{0.01\textwidth} \vspace{-3.5cm}
		\textbf{(a)}
	\end{subfigure}
    \includegraphics[width=0.35\textwidth]{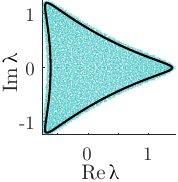}
    \begin{subfigure}[t]{0.01\textwidth} \vspace{-3.5cm}
		\textbf{(b)}
	\end{subfigure} \vspace{0.6cm}
     \includegraphics[width=0.35\textwidth]{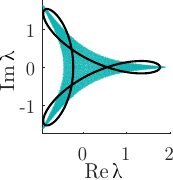}    
  \begin{subfigure}[t]{0.01\textwidth} \vspace{0.2cm} \hspace{-13.0cm}
		\textbf{(c)}
	\end{subfigure}\vspace{-0.4cm}
    \includegraphics[width=0.8\textwidth]{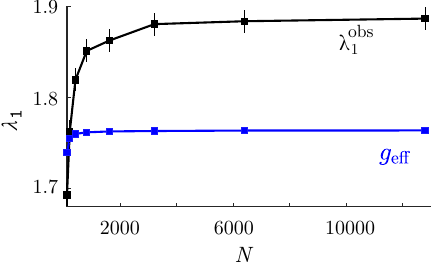} \vspace{0.0cm}
     \caption{(a) Eigenvalue distribution (cyan dots) shown for $\rho \approx 0.38$ with $N=6400$. The black curve shows the solution of the equation defining the support of the eigenvalue distribution $z(\phi)=e^{i \phi}+\rho e^{-2 i \phi}$. (b) The same as (a) but with $\rho \approx 0.76$.  (c) The observed values of the real part of the rightmost eigenvalue, $\lambda_1^{\text{obs}}$ (black) and $g_\mathrm{eff}$ (blue) as a function of $N$, with $g=1$ and $\rho=0.76$, averaged over 30 different realizations.}
    \label{fig:lambda1_vs_geff}
\end{figure}

\subsection{Numerical details of phase diagram}
\hypertarget{chaosdetails}{}\label{supp:numerical_methods}
The probability of observing chaotic, oscillatory, or fixed point behavior was computed by averaging the results obtained using $300$ different network realizations with $N=1600$. To determine a fixed-point solution we computed the absolute value of the numerically determined derivative of the individual trajectory, and tested if this value was smaller than $10^{-4}$ in the final $20 \%$ of the total runtime of the dynamics. If this test indicated that a fixed point was not reached, we estimate the Lyapunov exponents with a cutoff of $2 \times 10^{-3}$ to separate chaotic from oscillatory solutions. Colors were shaded based on their corresponding probabilities with red, green, and blue, for chaos, oscillations, and fixed points respectively.

\subsection{Probabilities of observing chaotic dynamics}
\hypertarget{Fitting_chaos_probabilities}{}\label{supp:Fitting chaos probabilities}
The probability of observing chaotic behavior as a function of \(g_{\mathrm{eff}}\) for various network sizes, \(N\), is shown in \hyperlink{FigureS3}{Fig.\ S3}. We classified dynamics with Lyapunov exponents exceeding \(2\times10^{-3}\) as chaotic. We used different numbers of realizations, \(N_R\): 300 for \(N\le2400\), 200 for \(N=3200\), 100 for \(N=4800\) and \(6400\), and 50 for \(N=12800\).
Each curve in \hyperlink{FigureS3}{Fig.\ S3} was fitted with
$P(g_{\mathrm{eff}}) = \frac{1}{2}\Bigl[1 + \tanh\bigl(b\,(g_{\mathrm{eff}} - c)\bigr)\Bigr]$.
From these fits, we extracted the value of \(g_{\mathrm{eff}}\) corresponding to \(P=0.5\).

The extracted values of $g_{\mathrm{eff}}$ as a function of $1/\sqrt{N}$ are plotted in \hyperlink{FigureS4}{Fig.\ S4}. Each set of data points was fit to a linear model with  $g_{\mathrm{eff}}=a+b N^{-1/2}$. \hyperlink{FigureS4}{Figure\ S4(a)-S4(b)} shows the data corresponding to $\rho=0$ and $\rho \approx 0.23$ respectively. The fitted curves in both cases show excellent agreement with a linear model ($R^2\approx 0.99$) with an interception point slightly above $g_{\mathrm{eff}}=1$. \hyperlink{FigureS4}{Figure\ S4(c)} depicts the data corresponding to $\rho \approx 0.76$. A linear fit fails to describe the data, so we fit the data to a model of the form $g_{\mathrm{eff}}=a+b N^{-c/2}$. The extrapolated interception point is far from $1$, showing a significant qualitative change compared with $\rho \approx 0$ and $\rho \approx 0.23$. These results support our hypothesis that chaotic behavior is suppressed at the onset of instability of the origin when correlations are strong.

  \setlength{\textfloatsep}{10pt plus 1.0pt minus 2.0pt}

\begin{figure}\hypertarget{FigureS3}{}
    \centering 
\begin{subfigure}[t]{0.01\textwidth} \vspace{-4.8cm}
		\textbf{(a)}
	\end{subfigure} \vspace{0.6cm}
    \includegraphics[width=0.7\textwidth]{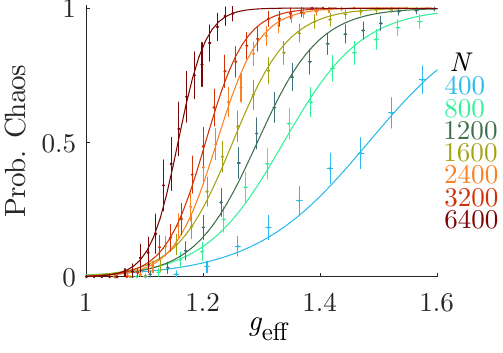} 

\begin{subfigure}[t]{0.01\textwidth} \vspace{-4.8cm}
		\textbf{(b)}
	\end{subfigure}\vspace{0.6cm}
    \includegraphics[width=0.7\textwidth]{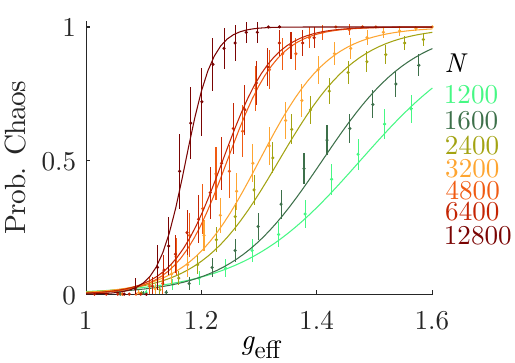}

    \begin{subfigure}[t]{0.01\textwidth} \vspace{-4.8cm}
		\textbf{(c)}
	\end{subfigure}
     \includegraphics[width=0.7\textwidth]{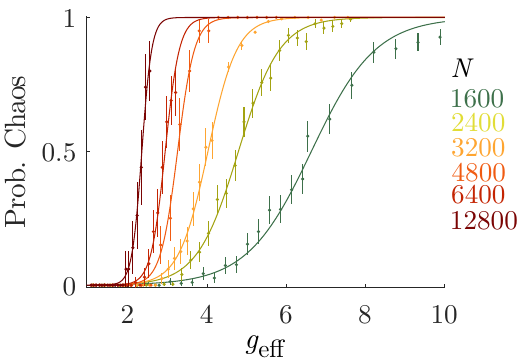}
 \vspace{-0.0cm}
    \caption{Probability of observing chaotic behavior as a function of $g_{\mathrm{eff}}$ for different values of $N$ with  (a) $\rho \approx 0$, (b) $\rho \approx 0.23$, and (c) $\rho \approx 0.76$. Error bars depict $2 \sigma$ errors, solid lines show the fit $P(g_{\mathrm{eff}})=\tfrac12\bigl[1+\tanh\bigl(b(g_{\mathrm{eff}}-c)\bigr)\bigr]$.}
    \label{fig:rho_023_N}
\end{figure} 

In theory, as $N\rightarrow \infty$ and $P\rightarrow1$, the expected intercept in \hyperlink{FigureS4}{Fig.\ S4(a)} is $g_{\mathrm{eff}}\rightarrow1$. Thus, finite-size effects and the choice of $P=0.5$ as our reference probability point may cause small deviations. The value $P=0.5$ was chosen to avoid any numerical issues that may arise when $P$ approaches $1$. Additionally, in \hyperlink{FigureS3}{Fig.\ S3(c)}, the probabilities of observing chaotic behavior provide a slight overestimate of the corresponding probabilities in large networks. Independent runs reveal that even apparently highly chaotic dynamics can suddenly converge to a limit cycle or fixed point. This transient behavior becomes increasingly difficult to detect as $N$ grows. Consequently, the values of $g_{\mathrm{eff}}$ in this panel should be treated as a lower bound for the true values, making chaotic behavior at large $N$ even less probable than reported here.

\begin{figure}\hypertarget{FigureS4}{}
    \centering 
\begin{subfigure}[t]{0.01\textwidth} \vspace{-4.8cm}
		\textbf{(a)}
	\end{subfigure}\vspace{0.6cm}
    \includegraphics[width=0.7\textwidth,trim=-0.4cm 0pt 0.0cm 0.0cm,clip]{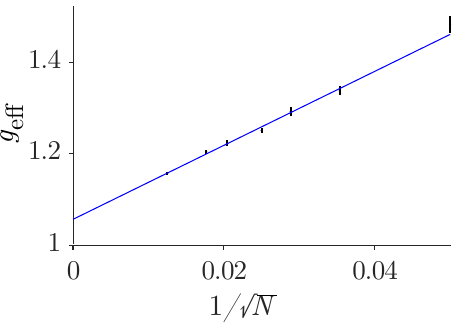} 

\begin{subfigure}[t]{0.01\textwidth} \vspace{-4.8cm}
		\textbf{(b)}
	\end{subfigure}\vspace{0.6cm}
    \includegraphics[width=0.7\textwidth,trim=-0.4cm 0pt 0.0cm 0.0cm,clip]{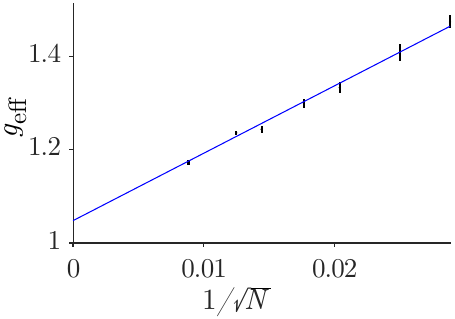} 
        
    \begin{subfigure}[t]{0.01\textwidth} \vspace{-4.8cm}    
		\textbf{(c)}
	\end{subfigure}
     \includegraphics[width=0.7\textwidth,trim=-0.4cm 0pt 0.0cm 0.0cm,clip]{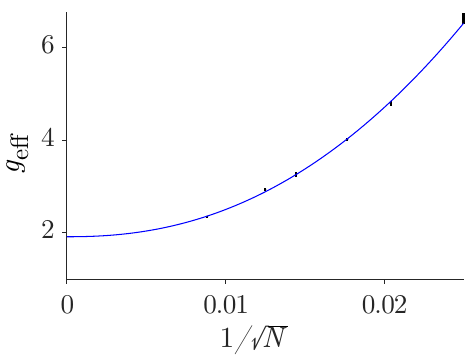}
 \vspace{-0.0cm}
    \caption{Black points show $g_{\mathrm{eff}}$ as a function of $1/
    \sqrt{N}$ for (a) $\rho \approx 0$, (b) $\rho \approx 0.23$, and (c) $\rho \approx 0.76$ with $2 \sigma$ error bars. The blue solid lines depict a linear fit $y = a + bx$ in (a) and (b), and a power-law fit $y = a + b\,x^c$ in (c). Fit parameters: (a) $a=1.06\pm 0.01$,   $b=8.1\pm 0.3$; (b) $a=1.05\pm 0.01$,   $b=14.5\pm 0.7$;  (c) $a=1.9\pm 0.7, \  b=(1.83\pm1.77) \times 10^{4},\ c=2.2\pm0.3.$ }
\end{figure}

\subsection{Error estimation and fitting process}
\hypertarget{Error estimation and fitting process}{}\label{supp:Error estimation and fitting process}
We first fit each set of data points in \hyperlink{FigureS3}{Fig.\ S3} to a sigmoid function of the form $P(g_{\mathrm{eff}})=\tfrac12\bigl[1+\tanh\bigl(b(g_{\mathrm{eff}}-c)\bigr)\bigr]$. For each chosen value of  $g_{\mathrm{eff}}$, there is a true value of the rightmost eigenvalue, $\lambda$. Thus, henceforth, we refer to $g_{\mathrm{eff}}$ as the mean real part of the leading eigenvalue $\mbox{Re}\; \lambda$,  $\bar\lambda$, with a standard error $\sigma_{\bar\lambda}=s/\sqrt{N_R}$ from $N_{R}$ realizations (with sample standard deviation $s$).  The probability of observing chaotic dynamics, $P$, has a binomial standard error $\sigma_P=\sqrt{P(1-P)/N_{R}}$.
The parameters $b$ and $c$ are obtained by weighted nonlinear least squares with inverse-variance weighting. 
To account for uncertainty in both axes, we use an effective variance $\sigma_{\rm eff}^2 = \sigma_p^2 + \bigl[f'(g_{\mathrm{eff}})\,\sigma_{\bar\lambda}\bigr]^2,$ with $f'(g_{\mathrm{eff}}) = \frac{b_0}{2}\sech^2 \!\bigl(b_0(g_{\mathrm{eff}}-c_0)\bigr)$, and weight each point with $1/\sigma_{\rm eff}^2$.
In practice, we implement this in two steps: A first fit using only $1/\sigma_{P}^2$ to obtain provisional $(b_0, c_0)$, followed by computing $f'(\bar{\lambda})$ at $(b_0, c_0)$ to obtain $\sigma_{\rm eff}^2$. We then refit the data with $1/\sigma_{\rm eff}^2$. The fit yields $b$, $c$ and the corresponding covariance matrix $\mathrm{Cov}(b,c)$. We note that the horizontal error, $\sigma_{\bar\lambda}$ is between one to two orders of magnitude smaller than the vertical error in the probability of observing chaotic behavior, thus, practically negligible in the fitting process. 

\begin{figure}\hypertarget{FigureS5}{}
    \centering
    \includegraphics[width=0.65\textwidth]{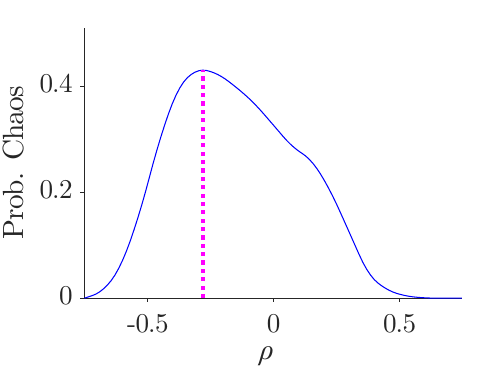}
    \caption{Probability of observing chaotic behavior as a function of $\rho$ for $g_{\mathrm{eff}}=1.2$. The magenta dotted line depicts the value of the maximum probability at $\rho \approx -0.28$.}
\end{figure}

Once $b$ and $c$ are estimated, we invert the sigmoid at a target probability $P_0$,
$\bar\lambda_{P_0}
= c + \frac{1}{b}\,\operatorname{arctanh}(2P_0-1),$
and propagate uncertainties via
$\sigma_{\bar\lambda_{P_0}}^2
= \Bigl(\frac{\partial\bar\lambda_{P_0}}{\partial b}\Bigr)^{2}\sigma_b^2
+ \Bigl(\frac{\partial\bar\lambda_{P_0}}{\partial c}\Bigr)^{2}\sigma_c^2
+ 2\,\frac{\partial\bar\lambda_{P_0}}{\partial b}\,\frac{\partial\bar\lambda_{P_0}}{\partial c}\,\mathrm{Cov}(b,c).$
Finally, we fit $\bar\lambda_{P_0}$ \textit{vs.}\ $1/\sqrt{N}$ to a linear model $y = a + b\,x$, or a power law model $y = a + b\,x^c$
using weights $1/\sigma_{\bar\lambda_{P_0}}^2$, and report the fitted parameters with their 2\,$\sigma$ uncertainties,  \hyperlink{FigureS4}{Fig.\ S4}. In this figure we show the results for $P_0=0.5$, where $\bar\lambda_{P_0}$ was replaced with $g_{\mathrm{eff}}$.

\begin{figure}[H]\hypertarget{FigureS6}{} \vspace{1.0cm}
    \centering 
\begin{subfigure}[t]{0.01\textwidth} \vspace{-5.0cm}
		\textbf{(a)}
	\end{subfigure} \vspace{0.5cm}
    \includegraphics[width=0.65\textwidth,trim=-0.4cm 0pt 0.0cm 0.0cm,clip]{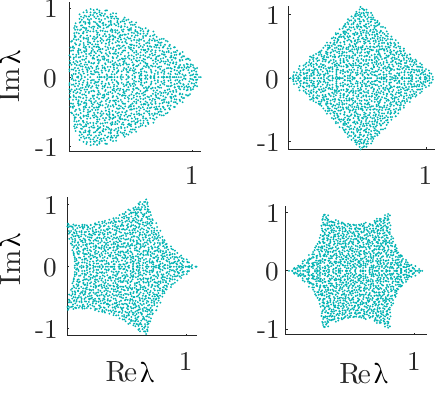} 

\begin{subfigure}[t]{0.01\textwidth} \vspace{-5.0cm}
		\textbf{(b)}
	\end{subfigure}
    \includegraphics[width=0.65\textwidth,trim=-0.4cm 0pt 0.0cm 0.0cm,clip]{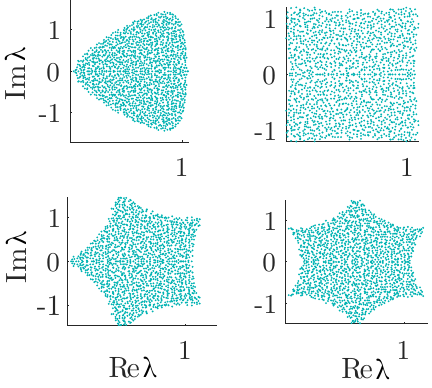} 
    \caption{Eigenvalue distributions for $\alpha=3$ (top left), $\alpha=4$ (top right), $\alpha=5$ (bottom left), and $\alpha=6$ (bottom right) with (a) $\rho\approx 0.18$, and (b) $\rho \approx -0.18$. Network size is $N=1600$ and $g_\mathrm{eff} \approx 1.1$.}
\end{figure} 








\end{document}